\documentclass[apjl]{emulateapj}

\usepackage{amsfonts,amsmath,graphicx,natbib,apjfonts,xcolor,subfigure,gensymb,longtable,ulem}

\def\swift{{\it Swift}}

\def\fr{1}
\def\icr{2}
\def\ciera{3}
\def\hsca{4}

\shorttitle{SN2014ad}
\shortauthors{Marongiu~et~al.}

\begin{document}
\title{Constraints on the environment and energetics of the broad-line Ic SN\,2014ad from deep radio and X-ray observations}

\author{M.~Marongiu\altaffilmark{\fr,\icr}
, C.~Guidorzi\altaffilmark{\fr}
, R.~Margutti\altaffilmark{\ciera}
, D. L.~Coppejans\altaffilmark{\ciera}
, R.~Martone\altaffilmark{\fr,\icr}
, A.~Kamble\altaffilmark{\hsca}
}

\altaffiltext{\fr}{Department of Physics and Earth Science, University of Ferrara, via Saragat 1, I--44122, Ferrara, Italy}
\altaffiltext{\icr}{ICRANet, Piazzale della Repubblica 10, 65122, Pescara, Italy}
\altaffiltext{\ciera}{Center for Interdisciplinary Exploration and Research in Astrophysics (CIERA) and Department of Physics and Astronomy, Northwestern University, Evanston, IL 60208, USA}
\altaffiltext{\hsca}{formerly at Harvard-Smithsonian Center for Astrophysics, 60 Garden St. Cambridge. MA 02138, USA}

\email{marco.marongiu@unife.it}

\begin{abstract}
Broad-line type Ic Supernovae (BL-Ic SNe) are characterized by high ejecta velocity ($\ga 10^4$~km s$^{-1}$) and are sometimes associated with the relativistic jets typical of long duration ($\ga 2$~s) Gamma-Ray Bursts (L-GRBs).
The reason why a small fraction of BL-Ic SNe harbor relativistic jets is not known.
Here we present deep X-ray and radio observations of the BL-Ic SN\,2014ad extending from $13$ to $930$ days post explosion.
SN\,2014ad was not detected at either frequency and has no observational evidence of a GRB counterpart.
The proximity of SN\,2014ad ($d\sim 26$~Mpc) enables very deep constraints on the progenitor mass-loss rate $\dot{M}$ and on the total energy of the fast ejecta $E$.
We consider two synchrotron emission scenarios for a wind-like circumstellar medium (CSM): (i) uncollimated non-relativistic ejecta, and (ii) off-axis relativistic jet.
Within the first scenario our observations are consistent with GRB-less BL-Ic SNe characterized by a modest energy budget of their fast ejecta ($E \la 10^{45}$~erg), like SNe 2002ap and 2010ay.
For jetted explosions, we cannot rule out a GRB with $E \la 10^{51}$~erg (beam-corrected) with a narrow opening angle ($\theta_j \sim 5^{\circ}$) observed moderately off-axis ($\theta_{\rm obs} \ga 30^{\circ}$) and expanding in a very low CSM density ($\dot{M}$ $\la 10^{-6}$~M$_{\odot}$ yr$^{-1}$).
Our study shows that off-axis low-energy jets expanding in a low-density medium cannot be ruled out even in the most nearby BL-Ic SNe with extensive deep observations, and might be a common feature of BL-Ic SNe.
\end{abstract}


%%%%%%%%%%%%%%%%%%%%%%%%%%%%%%%%%%%%%%%%%%%
\section{Introduction}
\label{sec:intro}
%%%%%%%%%%%%%%%%%%%%%%%%%%%%%%%%%%%%%%%%%%%
Type Ic supernovae (SNe) are hydrogen-stripped core-collapse explosions (CCSNe) of massive stars with M$_{ZAMS} \ga 8 M_{\odot}$ that show no evidence for hydrogen and helium in their spectra \citep{Filippenko97}.
Potential candidates for type Ic SN progenitors are massive Wolf Rayet (WR) stars and stars in close binary systems (\citealt{Ensman88,Galyam17}).
At the time of writing the exact nature of their progenitors is unclear
\citep{Podsiadlowsky92,Yoon10,Eldridge13,Smartt09,Smartt15,Dessart15a,Dessart15b,Dessart17}.
Notable in this respect is the recent detection of the progenitor system of the Ic SN\,2017ein \citep{Kilpatrick18,VanDyk18} which pointed to a massive stellar progenitor with $M \sim 60$~$M_{\odot}$ in a binary system.

Ic SNe typically show a bell-shaped radio spectrum powered by synchrotron emission and extending all the way to the X-ray band. The spectral peak frequency describes the transition between the optically thick part of the spectrum --below which synchrotron self-absorption (SSA) takes place-- and the optically thin portion of the spectrum \citep{RybickiLightman79,Chevalier98,ChevalierFransson06}.
The synchrotron emission is produced by electrons that are accelerated at the shock front between the SN ejecta and the the circumstellar medium (CSM). As the shock wave expands, the optical depth to SSA decreases and hence the spectral peak frequency cascades down to lower frequencies with time.
In a SN explosion, the X-ray  and radio emission resulting from the SN shock propagation in the medium track the fastest material ejected by the explosion, while the optical emission is of thermal origin and originates from the inner ejecta layers.

A small fraction ($\sim 4 \%$; \citealt{Shivvers17}) of Ic SNe, called broad-line Ic SNe (BL-Ic SNe), are characterized by broad lines in the optical spectrum implying large expansion velocities of the ejecta ($\ga 2\times 10^{4}$ km s$^{-1}$, e.g. \citealt{Mazzali02,Cano17rev}), $\sim 10^4$ km s$^{-1}$ faster than in ``ordinary'' Ic SNe \citep{Modjaz16}.
Some BL-Ic SNe are associated with ultra-relativistic jets that generate long duration ($\ga 2$ s) Gamma-Ray Bursts (L-GRBs, e.g. \citealt{Cano17rev}), which are observable at cosmological distances up to $z\sim10$ (e.g. \citealt{Cucchiara11b}).
In the local universe ($z\le0.1$) some BL-Ic SNe  have also been found in association with mildly relativistic outflows in low-luminosity GRBs (ll-GRBs, which are too weak to be detected at larger distances, \citealt{Liang07}).
As opposed to L-GRBs, ll-GRBs show no evidence for collimation of their fastest ejecta, i.e. no jet \citep{Kulkarni98,Soderberg06d,Bromberg11c}.

A possible interpretation of the observational lack of evidence for  L-GRB counterparts in the majority of BL-Ic SNe is the off-axis jet scenario \citep{Rhoads99,Eichler99,Yamazaki03,Piran04_rev,Soderberg06e,Bietenholz14VLBI,Corsi16}, where the explosion powers a GRB-like jet that  is misaligned with respect to the observer line of sight.
In this scenario, as the jet velocity gradually decreases and relativistic beaming becomes less severe, the emission becomes observable from increasingly larger viewing angles.
Deep radio and X-ray observations extending to hundreds of days post explosion offer the opportunity to reveal the emission from off-axis jets as well as to recover weak GRBs that would not trigger current $\gamma$-ray observing facilities.

\begin{figure} %[htbp] 
\centering
\includegraphics[width=8.9cm]{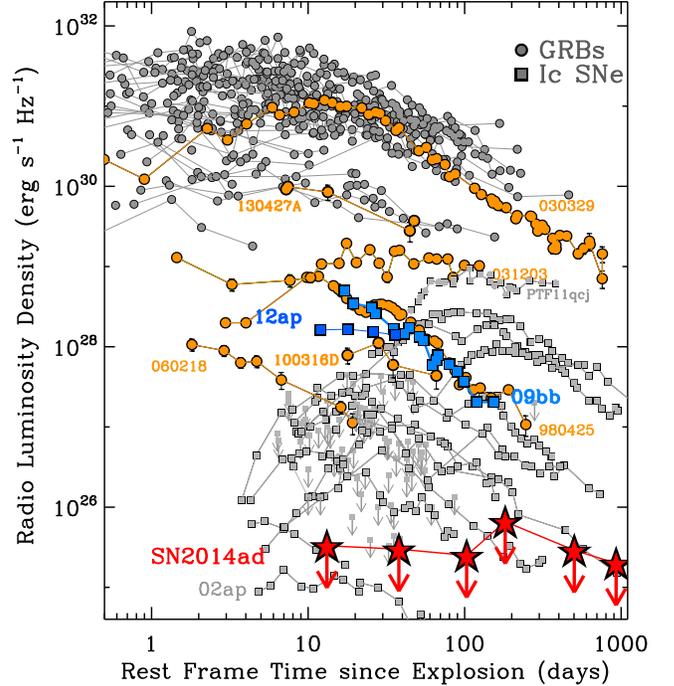}
\caption{Deep radio limits on the emission from SN\,2014ad (red stars) in the context of L-GRBs (circles; gray is for cosmological GRBs, while orange is for GRBs at $z\le 0.3$) and H-stripped CCSNe (squares; gray is for normal SNe, blue is for SNe with relativistic ejecta).
%Adapted from \cite{Margutti14b}.
Our deep radio limits on the emission from the BL-Ic SN\,2014ad are consistent with a luminosity comparable to that of SN\,2002ap. The detected radio emission from SN\,2002ap points to a non-relativistic (shock velocity $\sim 0.3c$) uncollimated explosion with a small energy budget of the fast ejecta ($E \sim 1.5 \times 10^{45}$~erg; \citealt{Berger02}). \\
}
\label{fig:radioworld}
\end{figure}

Here we present extensive ($\delta t\sim10-1000$ days) broad-band (radio to X-ray) observations  of SN\,2014ad, a BL-Ic SN that exploded in the galaxy PGC\,37625 (Mrk 1309) at $d = 26.44$~Mpc \citep{Sahu18}. SN\,2014ad is among the closest BL-Ic SNe  discovered to date, which enables very deep limits on its radio and X-ray emission (Figure~\ref{fig:radioworld} and Table~\ref{tab:tablevla1}).
We present constraints on the progenitor mass-loss rate $\dot{M}$ and the total energy of the fast ejecta $E$ in two scenarios: (i) mildly relativistic, nearly isotropic, synchrotron self-absorbed radio emission due to the SN ejecta ploughing through a wind-like CSM; (ii) synchrotron emission from a relativistic off-axis GRB-like jet.

The analysis of the optical emission from SN\,2014ad by \cite{Sahu18} and \cite{Stevance17} revealed that the bulk of its ejecta velocity is $\sim 3 \times 10^{4}$ km s$^{-1}$ at early times, with kinetic energy $E_k \sim (1.0 \pm 0.3) \times 10^{52}$~erg, larger than in type-Ic SNe, and similar to BL-Ic SNe and GRB-SNe.
The metallicity of the host galaxy of SN\,2014ad is $\sim 0.5$ Z$_{\odot}$.
The total explosion ejecta mass inferred by \cite{Sahu18} and \cite{Stevance17} is $M_{ej} \sim (3.3 \pm 0.8)$~M$_{\odot}$ suggesting a massive progenitor star with $M_{ZAMS} \ga 20$~M$_{\odot}$.
Spectropolarimetry by \citet{Stevance17} also suggests a mild deviation from a spherical geometry of the ejecta.

This paper is organized as follows.
In Section~\ref{sec:obs} we describe our radio and X--ray observations; in Section~\ref{sec:IC} we present the constraints on the environment derived from our X-ray limits, whereas in Section~\ref{sec:mod} we present environment constraints derived from the radio and X-ray broadband modeling in two different scenarios (i.e., an ``ordinary'' isotropic SN outflow, and a beamed relativistic jet).
Our results and analysis are discussed in Section~\ref{sec:disc} and conclusions are drawn in Section~\ref{sec:conc}.

%---------------------------------------
\section{Observations}
\label{sec:obs}
%---------------------------------------

SN\,2014ad was discovered by \citet{howerton14} on March $12.4$, 2014 (MJD $56728.4$) in public images from the Catalina Real-Time Transient Survey \citep{CATALINA} at $\alpha=11^{\rm h}57^{\rm m}44^{\rm s}.44$, $\delta=-10^{\circ}10'15.7''$.
Throughout this paper we assume a SN explosion date $t_0=56725\pm 3$~MJD \citep{Sahu18}; times given are in reference to this explosion date unless otherwise noted.

%---------------------------------------
\subsection{Radio Observations with the VLA}
\label{subsec:radiofu}
%---------------------------------------
VLA follow-up observations were carried out between March 22, 2014 (MJD 56738) and September 23, 2016 (MJD 57654), from $\sim 13$~d to $\sim 930$~d post explosion, under Proposal VLA/14A-531 (PI: Kamble).
Data were taken in eight spectral windows at L-band (with baseband central frequencies of $1.3$ and $1.7$~GHz, respectively), C-band ($5$ and $7$~GHz), X-band ($8.5$ and $11$~GHz), Ku-band ($13.5$ and $16$~GHz), with a nominal bandwidth of $\sim 1$~GHz ($\sim 0.4$~GHz for L-band).
3C286 and J1330-1449 were used as flux/bandpass and phase/amplitude calibrators, respectively.
The Common Astronomy Software Application ({\sc casa}, v. 4.7.2, \citealt{CASA})\footnote{\url{https://casa.nrao.edu/}} was used to calibrate, flag and image the data.
Images were formed from the visibility data using the CLEAN algorithm \citep{Hogbom74}.
The image size was set to ($1024 \times 1024$) pixels, the pixel size was determined as $1/5$ of the nominal beam width and the images were cleaned using natural weighting.
The upper limits on the flux densities were calculated at a $3 \sigma$ confidence level (Table~\ref{tab:tablevla1}).

\begin{deluxetable*}{cc|cccccccc}
\tablecolumns{9}
\tablecaption{Log of VLA observations of SN\,2014ad: observation central time $t_{mid}$, epoch $t_{\rm e} = t_{\rm mid}-t_0$ since the estimated explosion date $t_0$, VLA array configuration, beam size $\theta_{\rm FWHM}$, central frequency $\nu_c$ and its bandwidth $\Delta \nu$, the uncertainty $\sigma_S$, the upper limit on the flux density $S$ (at $3$--$\sigma$) and the relative luminosity $L_{25}$ (in units of $10^{25}$ erg s$^{-1}$ Hz$^{-1}$) of the source. In no case was the source was detected with $\ge 3$--$\sigma$ confidence.\label{tab:tablevla1}}
\tablehead{
\colhead{$t_{mid}$} & \colhead{$t_{\rm e}$} & \colhead{VLA} & \colhead{$\theta_{\rm FWHM}$} & \colhead{$\nu_c$} & \colhead{$\Delta \nu$} & \colhead{$\sigma_S$} & \colhead{S(3--$\sigma$)} & \colhead{$L_{25}$} \\
\colhead{[MJD]} & \colhead{[days]} & \colhead{Configuration} & \colhead{[arcsec]} & \colhead{[GHz]} & \colhead{[GHz]} & \colhead{[$\mu$Jy]} & \colhead{[$\mu$Jy]} & \colhead{[erg s$^{-1}$ Hz$^{-1}$]}
} %\colnumbers
\startdata
56738.19  &  13.19 & A   & 1.42 & 1.26 & 0.384 & 28.8 & 86.4 & 12.1 \\
          &        & A   & 0.93 & 1.80 & 0.384 & 30.8 & 92.4 & 12.9 \\
          &        & A   & 0.34 & 5.0  & 0.896 &  9.0 & 27.0 &  3.8 \\
          &        & A   & 0.24 & 7.1  & 0.896 &  8.1 & 24.3 &  3.4 \\
          &        & A   & 0.19 & 8.6  & 0.896 &  7.9 & 20.7 &  3.3 \\
          &        & A   & 0.15 & 11.0 & 0.896 &  7.8 & 23.4 &  3.3 \\
          &        & A   & 0.13 & 13.5 & 0.896 &  7.7 & 23.1 &  3.2 \\
          &        & A   & 0.11 & 16.0 & 0.896 &  9.1 & 27.3 &  3.8 \\
56763.21  &  38.21 & A   & 1.42 & 1.26 & 0.384 & 31.6 & 94.8 & 13.2 \\
          &        & A   & 0.93 & 1.80 & 0.384 & 31.4 & 94.2 & 13.1 \\
          &        & A   & 0.34 & 5.0  & 0.896 & 10.5 & 31.5 &  4.4 \\
          &        & A   & 0.24 & 7.1  & 0.896 &  7.2 & 21.6 &  3.0 \\
          &        & A   & 0.19 & 8.6  & 0.896 &  8.0 & 24.0 &  3.4 \\
          &        & A   & 0.15 & 11.0 & 0.896 & 10.3 & 30.9 &  4.3 \\
          &        & A   & 0.13 & 13.5 & 0.896 &  7.3 & 21.9 &  3.1 \\
          &        & A   & 0.11 & 16.0 & 0.896 &  7.6 & 22.8 &  3.2 \\
56828.96  & 103.96 & AnD &12.02 & 5.0  & 0.896 &  6.9 & 20.7 &  2.9 \\
          &        & AnD & 7.93 & 7.1  & 0.896 &  5.2 & 15.6 &  2.2 \\
          &        & AnD & 0.20 & 8.6  & 0.896 &  6.0 & 18.0 &  2.5 \\
          &        & AnD & 0.15 & 11.0 & 0.896 &  6.3 & 18.9 &  2.6 \\
56906.76  & 181.76 & D   &12.02 & 5.0  & 0.896 & 13.9 & 41.7 &  5.8 \\
          &        & D   & 7.93 & 7.1  & 0.896 & 10.8 & 32.4 &  4.5 \\
          &        & D   & 6.98 & 8.6  & 0.896 & 15.7 & 47.1 &  6.6 \\
          &        & D   & 5.46 & 11.0 & 0.896 & 16.8 & 50.4 &  7.0 \\
57227.81  & 502.81 & A   & 0.34 & 5.0  & 0.896 &  9.9 & 26.7 &  4.1 \\
          &        & A   & 0.24 & 7.1  & 0.896 &  9.4 & 27.7 &  3.9 \\
          &        & A   & 0.19 & 8.6  & 0.896 &  6.9 & 20.7 &  2.9 \\
          &        & A   & 0.15 & 11.0 & 0.896 & 10.1 & 30.3 &  4.2 \\
57654.66  & 929.66 & B   & 1.15 & 5.0  & 0.896 &  6.6 & 19.8 &  2.8 \\
          &        & B   & 0.79 & 7.1  & 0.896 &  6.3 & 18.9 &  2.6 \\
          &        & B   & 0.65 & 8.6  & 0.896 &  7.0 & 21.0 &  2.9 \\
          &        & B   & 0.51 & 11.0 & 0.896 &  6.8 & 20.4 &  2.8 
          \enddata
\end{deluxetable*}

%---------------------------------------
\subsection{X--ray Observations with Swift-XRT}
\label{subsec:XRT}
%---------------------------------------

The X-Ray Telescope (XRT; \citealt{Burrows05}) onboard the \swift{} Gehrels spacecraft \citep{Gehrels04} observed the region of SN\,2014ad in Photon Counting (PC) mode several times from March 19, 2014 to March 11, 2017.
We find no evidence for statistically significant X-ray emission at the location of SN\,2014ad.
We extracted the $0.3$--$10$~keV light curve, consisting of $3\sigma$ upper limits, using the web interface provided by Leicester University\footnote{\url{http://www.swift.ac.uk/user\_objects/}}, which used {\sc heasoft} (v. 6.22).
We performed flux calibration by assuming an absorbed simple power-law spectral model ({\sc wabs*powerlaw} within {\sc xspec}) with column density frozen to the Galactic value along the SN line of sight, $N_{H,{\rm Gal}}=3.1\times10^{20}$~cm$^{-2}$ \citep{Kalberla05}.
We assumed a conservative value for the photon index, $\Gamma=2$, and derived the upper limit to the flux density at $1$~keV.
Finally, we calculated three light curves with different integration times: $10^5$, $2\times10^5$, and $5\times10^5$~s, respectively. Table~\ref{tab:tablexrt} reports the values for the longest timescale having the deepest limits.
We also calculated the corresponding $3\sigma$ upper limits on the $0.3$--$10$~keV luminosity.

\begin{deluxetable}{lrccr}
\tablecolumns{5}
\tablewidth{0pc}
\tablecaption{\swift-XRT 3--$\sigma$ upper limits on the flux density at $1$~keV ($F_{\nu,\;{\rm 1~KeV}}$) and $0.3$--$10$~keV luminosity ($L_{0.3-10}$). $t_{\rm e} = t_{\rm mid}-t_0$ is the epoch since the estimated SN explosion date $t_0$, $\Delta t$ is the bin time.
\label{tab:tablexrt}}
\tablehead{
\colhead{$t_{\rm mid}$} & \colhead{$t_{\rm e}$} & \colhead{$\Delta t$} & \colhead{$F_{\nu,\;{\rm 1~KeV}}$} & \colhead{$L_{0.3-10}$}\\
\colhead{[MJD]} & \colhead{[days]} & \colhead{[days]} & \colhead{[$\mu$Jy]} & \colhead{[erg\,s$^{-1}$]}
}
\startdata
56738.1 & $13.1$ & $5.8$ & $< 1.3\times10^{-2}$ & $< 1.0\times10^{42}$\\
56743.9 & $18.9$ & $5.8$ & $< 1.2\times10^{-2}$ & $< 9.0\times10^{41}$\\
56749.6 & $24.6$ & $5.8$ & $< 1.7\times10^{-2}$ & $< 1.3\times10^{42}$\\
56755.4 & $30.4$ & $5.8$ & $< 4.1\times10^{-2}$ & $< 3.2\times10^{42}$\\
57774.0 & $1049.0$ & $5.8$ & $< 0.11$ & $< 8.5\times10^{42}$\\
57808.7 & $1083.7$ & $5.8$ & $< 1.1$ & $< 8.5\times10^{43}$\\
57820.2 & $1095.2$ & $5.8$ & $< 6.7\times10^{-2}$ & $< 5.2\times10^{42}$\\
\enddata
\end{deluxetable}

%-------------------------------------------
\section{Constraints on the environment density from inverse Compton emission}
\label{sec:IC}
%-------------------------------------------

Inverse Compton (IC) emission from the upscattering of optical photospheric photons into the X-ray band by relativistic electrons at the shock front has been demonstrated to dominate the X-ray emission from H-stripped CCSNe that explode in low-density environments ($\dot{M} \la 10^{-5}$~M$_{\odot}$ yr$^{-1}$) at $\delta t\la 30$~d (e.g. \citealt{Bjornsson04}; \citealt{ChevalierFransson06}).
We adopt the IC formalism by \cite{Margutti12b} modified to account for the outer density structure of progenitors of BL-Ic SNe (which are likely to be compact) as in \cite{Margutti14b}.
The IC emission depends on: (i) the density structure of the SN ejecta and of the CSM; (ii) the electron distribution responsible for the up-scattering; (iii) explosion parameters (ejecta mass $M_{\rm{ej}}$ and kinetic energy\footnote{This is the kinetic energy carried by the slowly moving material powering the optical emission.} $E_{\rm{k}}$); and (iv) the bolometric luminosity of the SN: $L_{\rm{IC}}\propto L_{\rm{bol}}$.

For compact progenitors that are relevant here, the density scales as $\rho_{\rm{SN}}\propto r^{-n}$ with $n\sim10$ (see e.g. \citealt{Matzner99}; \citealt{ChevalierFransson06}).
We further assume a power-law electron distribution $n_{e}(\gamma)\propto \gamma^{-p}$ with $p\sim3$ as found in radio observations of type H-stripped CCSNe \citep{ChevalierFransson06} and a fraction of energy into relativistic electrons $\epsilon_e=0.1$.
We use the explosion parameters $E_{\rm{k}}=(1\pm0.3)\times 10^{52}$~erg and $M_{\rm{ej}}=(3.3\pm 0.8)$~M$_{\sun}$.
For a wind-like CSM structure $\rho_{\rm{CSM}}\propto r^{-2}$ with a typical wind velocity $v_w=1000$~km s$^{-1}$ as appropriate for massive stars (and hence BL-Ic SN progenitors, e.g. \citealt{Smith14}), the \swift{}-XRT non-detections at $\delta t<30$~d yield $\dot M<5\times10^{-5}$~M$_{\sun}$ yr$^{-1}$.

%-------------------------------------------
\section{Broadband modeling}
\label{sec:mod}
%-------------------------------------------
We interpret our deep radio and X-ray limits in the context of synchrotron self-absorbed (SSA) emission from either (i) uncollimated (i.e. spherical) non-relativistic ejecta (Sect.~\ref{subsec:ssamod}), or (ii) relativistic GRB-like jet (Sect.~\ref{sec:constraints}).

%-------------------------------------------
\subsection{SSA emission from  non-relativistic uncollimated ejecta}
\label{subsec:ssamod}
%-------------------------------------------
We follow \cite{Soderberg05} and adopt their formalism in the context of the radio emission from non-relativistic SN ejecta interacting with a wind-like CSM. The brightness temperature of a source is:
\begin{equation}
T_B\ =\ \frac{c^2}{2\pi k}\ \frac{f_\nu\,d^2}{(v_{\rm ph} t)^2\,\nu^2}\ \;,
\label{eq:t_b}
\end{equation}
where $c$ is the speed of light, $k$ is the Boltzmann constant, $f_\nu$ is the flux density at observed frequency $\nu$, $d$ is the source distance, $v_{\rm ph}$ is the photospheric velocity and $t$ is the observational epoch. For SN\,2014ad  we find $T_B \la 2.8 \times 10^{11}$~K at $t \sim 13.2$~d, where $v_{\rm ph} \sim 3.2 \times 10^4$~km s$^{-1}$ and $f_\nu < 86.4$~$\mu$Jy at $\nu=1.26$~GHz (Table \ref{tab:tablevla1}). Our inferred $T_B$ does not violate the $10^{12}$\,K limit of the Inverse Compton Catastrophe (ICC; \citealt{Kellermann81}), consistent with the expectations from a non relativistic spherical SSA source.

In the SSA model radiation originates from an expanding spherical shell of shock-accelerated electrons with radius $r$ and thickness $r/\eta$ (here we assume the standard scenario of a thin shell with $\eta = 10$; e.g. \citealt{LiChevalier99,Soderberg05}).
As the shock wave propagates through the CSM, it accelerates relativistic electrons into a power-law distribution $N(\gamma) \propto \gamma^{-p}$ for $\gamma\ge\gamma_m$, where $\gamma_m$ is the minimum Lorentz factor of the electrons \citep{Chevalier82,Chevalier98}. In this analysis we assume $p \sim 3$ as typically found in H-stripped core-collapse SNe (e.g. \citealt{ChevalierFransson06}).
The post-shock energy fraction in the electrons and magnetic field is given by $\epsilon_e$ and $\epsilon_B$, respectively; we further adopt equipartition of the post-shock energy density of the radio-emitting material between relativistic electrons and magnetic fields ($\epsilon_e = \epsilon_B = 1/3$).

The synchrotron emission from SNe typically peaks at radio frequencies on timescales of a few days to weeks after the SN explosion (e.g. \citealt{Corsi14}); this emission is suppressed at low frequencies by absorption processes.
\cite{Chevalier98} showed that the dominant absorption process is internal SSA for H-stripped SNe, and external free-free absorption (FFA) in H-rich SNe, as H-rich SNe tend to explode in higher density media.

Following \cite{Soderberg05}, the temporal evolution of the magnetic field $B(t)$, minimum Lorentz factor $\gamma_m(t)$, shock radius $r(t)$ and the ratio $\Im = \epsilon_e / \epsilon_B$) can be parametrized as:
\begin{equation}
B\ =\ B_0 \left(\frac{t-t_e}{t_0-t_e}\right)^{\alpha_B} \quad \gamma_m = \gamma_{m,0} \left(\frac{t-t_e}{t_0-t_e}\right)^{\alpha_{\gamma}}
\label{eq:rb}
\end{equation}
\begin{equation}
r\ =\ r_0 \left(\frac{t-t_e}{t_0-t_e}\right)^{\alpha_r} \quad \Im = \Im_0 \left(\frac{t-t_e}{t_0-t_e}\right)^{\alpha_{\Im}}
\label{eq:gf}
\end{equation}
where $r_0$, $B_0$, $\Im_0$ and $\gamma_{m,0}$ are measured at an arbitrary reference epoch $t_0$, and $t_e$ is the explosion time.
In this paper we adopt $t_0 = 13.2$~d (for which $r_0 \sim v_{ph} \times t_0 = 4 \times 10^{15}$~cm) and $t_e = 0$~d.
The temporal indices $\alpha_{r}$, $\alpha_{B}$, $\alpha_{\Im}$, $\alpha_{\gamma}$ are determined by the hydrodynamic evolution of the ejecta, as described in \citet{Soderberg05}.
In particular, $\alpha_{r}$ and $\alpha_{\Im}$ can be expressed as:
\begin{equation}
\alpha_r\ =\ \frac{n-3}{n-s},
\label{eq:alphar}
\end{equation}
\begin{equation}
\alpha_{\Im}\ =\ -s \alpha_r + \alpha_{\gamma} - 2\alpha_B,
\label{eq:alphaf}
\end{equation}
where $n$ and $s$ describe the density profile of the outer SN ejecta ($\rho_{ej} \propto r^{-n}$), and of the CSM ($\rho_{CSM} \propto r^{-s}$)\footnote{$s=0$ corresponds to the case of ISM-like CSM and $s=2$ correspond to the case of wind-like CSM.}, respectively.
The self-similar conditions $s < 3$ and $n > 5$ result in $\sim 0.5 < \alpha_r < 1$ \citep{Chevalier82}.
In this work we consider a wind-like CSM case (i.e. $s = 2$), and $n = 10$ as appropriate for massive compact stars that are thought to be progenitors of H-stripped CCSNe.
In the standard scenario \citep{Chevalier96}, $\epsilon_e$ and $\epsilon_B$ do not vary with time, from which we derive through Eq.~\ref{eq:gf} that $\alpha_{\Im}=0$, implying that:
\begin{equation}
\alpha_B\ =\ \left(\frac{2-s}{2}\right) \alpha_r - 1,
\label{eq:alphab_sm}
\end{equation}
\begin{equation}
\alpha_{\gamma}\ =\ 2\, (\alpha_r-1).
\label{eq:alphagamma_sm}
\end{equation}
Since $\alpha_{\Im} = 0$ and under the equipartition hypothesis ($\Im = 1$; Eq.~\ref{eq:gf}), it follows that $\alpha_r = 0.875$ (Eq.~\ref{eq:alphar}), $\alpha_B = -1$ (Eq.~\ref{eq:alphab_sm}) and $\alpha_{\gamma} = -0.25$ (Eq.~\ref{eq:alphagamma_sm}).

Under these assumptions and through Eq.~(\ref{eq:rb}), the characteristic synchrotron frequency is:
\begin{equation}
\begin{split}
\nu_m(t)\ =\ & \gamma_m^2 \frac{q B}{2 \pi m_e c}\ =\ \gamma_{m,0}^2 \frac{q B_0}{2 \pi m_e c}
\left(\frac{t}{t_0}\right)^{2\alpha_{\gamma}+\alpha_B} \\
& = \nu_{m,0} \left(\frac{t}{t_0}\right)^{2\alpha_{\gamma}+\alpha_B}\;,
\label{eq:num2}
\end{split}
\end{equation}
where $q$ is the electron charge and $m_e$ is the electron mass.
The frequency $\nu_{m,0}\equiv \nu_{m}(t_0)$ depends on $\gamma_{m,0}$ and $B_0$ as follows:
\begin{equation}
\nu_{m,0}\ =\ \gamma_{m,0}^2 \frac{q B_0}{2 \pi m_e c} \;.
\label{eq:num3}
\end{equation}

The radio flux density at a given observing frequency $\nu$ and epoch $t$ is thus given by:
\begin{equation}
\begin{split}
F(t,\nu)\ =\ & 10^{26}\ C_f \left(\frac{t}{t_0}\right)^{(4\alpha_r - \alpha_B)/2} \left(1 - e^{-\tau_{\nu}^{\xi}}\right)^{1/\xi} \\
& \times {\nu}^{5/2}\ F_3(x)\ F_2^{-1}(x)\ \ \ {\rm{mJy}}
\end{split}
\label{eq:denflux}
\end{equation}
with the optical depth $\tau_{\nu}$:
\begin{equation}
\tau_{\nu}(t)\ =\ C_{\tau} \left(\frac{t}{t_0}\right)^{\alpha_r + (3+p/2)\alpha_B + (p-2)\alpha_{\gamma} + \alpha_{\Im}} \nu^{-(p+4)/2}\ F_2(x)\;.
\label{eq:optdepth}
\end{equation}
$C_f$ and $C_{\tau}$ are normalization constants (see Appendix A2 of \citealt{Soderberg05}), $F_2(x)$ and $F_3(x)$ are Bessel functions with $x = 2/3\,(\nu/\nu_m)$, $\xi = [0, 1]$ describes the sharpness of the spectral break between optically thin and thick regimes.
We adopt $\xi = 1$.

As we can see from Eqs~(\ref{eq:denflux}, \ref{eq:optdepth}, \ref{eq:alphar} and \ref{eq:num2}), $F(t,\nu)$ depends on $C_f$, $C_{\tau}$, $p$, $n$, $s$, $\nu_{m,0}$, and $\xi$.
From Eqs~6--8 of \citet{Soderberg05} $C_f$ and $C_{\tau}$ can be expressed in terms of $r_0$, $B_0$ and $\eta$; thus, also using (\ref{eq:num3}), $F(t,\nu)$ can be expressed as a function of $r_0$, $B_0$, $p$, $n$, $s$, $\gamma_{m,0}$, $\eta$, and $\xi$, which are all fixed apart from $B_0$ and $\gamma_{m,0}$.
These two free parameters can be further expressed as a function of physically more useful quantities\footnote{These parameters are showed in Eqs.~13 and 14 of \citet{Soderberg05}, respectively.}, the SN progenitor mass-loss rate ($\dot{M}$) and the total kinetic energy of the radio-bright (fast) ejecta ($E$):
\begin{equation}
B_0 = \left(\frac{2 \eta \epsilon_e}{\Im_0 r_0^3}\right)^{1/2} \ E^{1/2} \;
\label{eq:b_0}
\end{equation}
\begin{equation}
\gamma_{m,0} = \Big(\frac{p-2}{p-1}\Big) \ \frac{2 m_p \epsilon_e v_w}{m_e c^2 r_0} \ \left(\frac{E}{\dot{M}}\right)
\label{eq:gamma_0}
\end{equation}
where $m_p$ is the proton mass and $v_w$ is the wind velocity.
Consequently, we express $\nu_{m,0}$ as a function of $\dot{M}$ and $E$ from (\ref{eq:num3}):
\begin{equation}
\nu_{m,0} = \left(\frac{p-2}{p-1}\right)^2 \frac{2 q}{\pi m_e c} \left(\frac{m_p v_w}{m_e c^2}\right)^2 \left(\frac{2 \eta \epsilon_e^5}{r_0^7 \Im_0}\right)^{1/2} \left(\frac{E^{5/2}}{\dot{M}^2}\right) \;.
\label{eq:nu_0b}
\end{equation}
As a result, $F(t,\nu)$ just depends on $\dot{M}$ and $E$.

\begin{figure} %[htbp] 
\centering
\includegraphics[width=8.8cm]{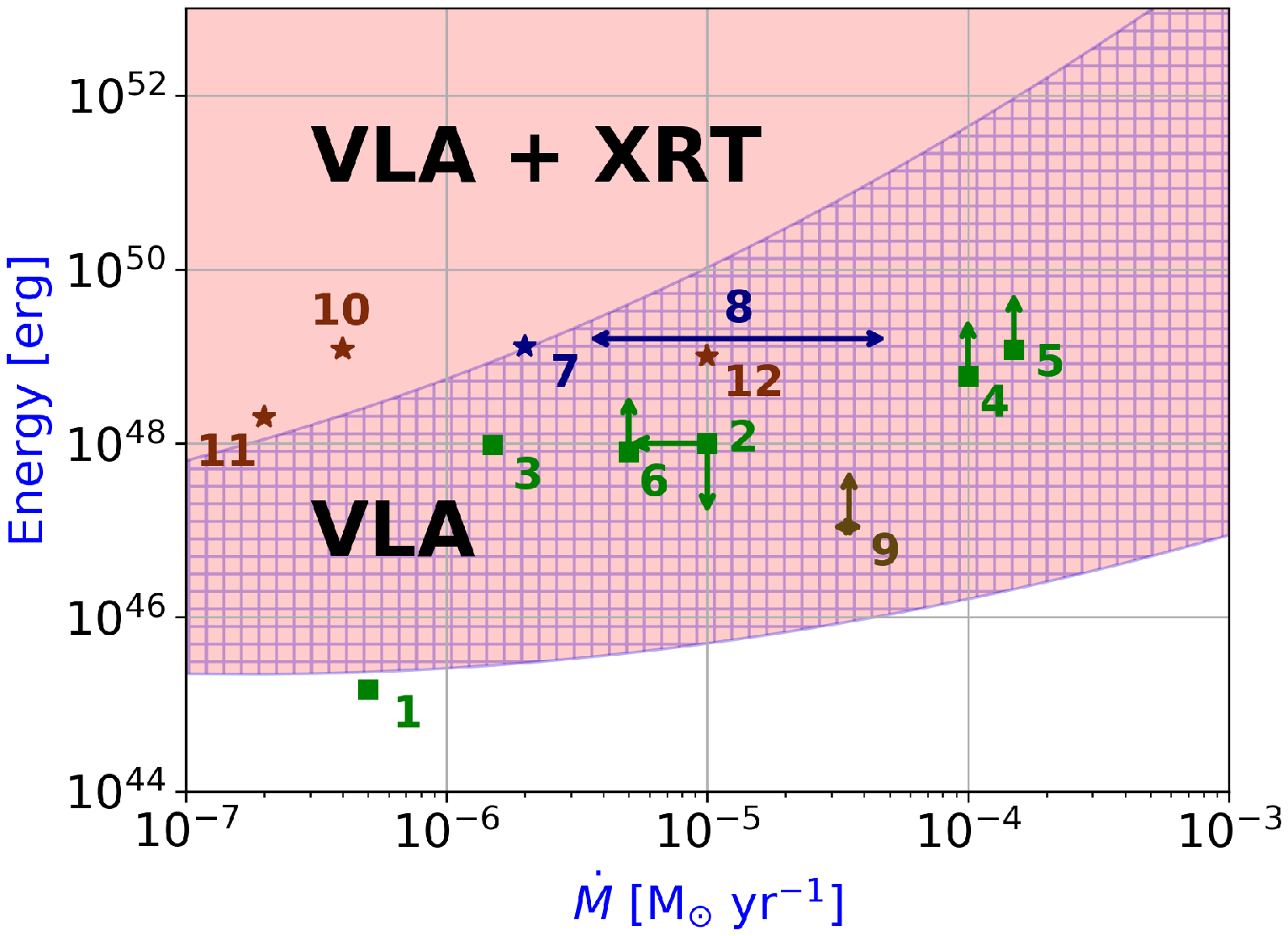}
\caption{
Regions of the total kinetic energy of the fast ejecta--mass-loss rate space excluded by VLA (hatched area) and VLA + XRT (red area) upper limits (see Table~\ref{tab:tablevla1}), as derived assuming the SSA model for a mildly relativistic, nearly isotropic explosion (Sect.~\ref{subsec:ssamod}).
In addition we show: some peculiar BL-Ic SNe (in green) [1] SN\,2002ap \citep{Berger02}, [2] SN\,2010ay \citep{Sanders12b}, [3] SN\,2007bg \citep{salas13}, [4-5-6] PTF\,11cmh - PTF\,11qcj - PTF\,14dby \citep{Corsi16}; the relativistic SNe (in blue) [7] SN\,2009bb \citep{Soderberg10a} and [8] SN\,2012ap \citep{Chakraborti15}; [9] SN\,2016coi (brown; \citealt{Terreran19}); the ll-GRBs (in red) [10] SN\,1998bw/GRB\,980425 \citep{LiChevalier99}, [11] SN\,2006aj/GRB\,060218 \citep{Soderberg06d} and [12] SN\,2010bh/GRB\,100316D \citep{Margutti13b}. \\
\label{fig:grid_nr_sn}
}
\end{figure}

We use a grid of $\dot{M}$ and $E$ values to compare our VLA upper limits (Table \ref{tab:tablevla1}) with the flux densities derived from (\ref{eq:denflux}).
In Figure~\ref{fig:grid_nr_sn} we explore the kinetic energy vs. mass-loss rate parameter space considering the (i) radio upper limits (hatched) and (ii) the radio limits plus the X-ray limits (red), which results in more stringent constraints: $E \la 10^{45}$~erg for $\dot{M}\la 10^{-6}~M_\odot$\,yr$^{-1}$ and $E \la 10^{46}$~erg for $\dot{M}\la 10^{-4}M_\odot$\,yr$^{-1}$.  We end by noting that at these low mass-loss rates the effects of FFA are negligible  (e.g. \citealt{Weiler86,FranssonBjornsson98}).

%-----------------------------------------------------
\subsection{SSA emission from a relativistic GRB-like jet}
\label{sec:constraints}
%-----------------------------------------------------

We generated a grid of radio light-curves powered by synchrotron emission from off-axis relativistic jets using the  {\sc boxfit} code (v2; \citealt{Vaneerten12b}), which is based on high-resolution, two-dimensional relativistic hydrodynamical simulations of relativistic jets. All the synthetic light curves were compared to our VLA upper limits (Table \ref{tab:tablevla1}) to determine the allowed region in the parameter space, using the same procedure as in \citet{Coppejans18}.

The radio emission from an off-axis jet depends on the following physical parameters: (1) isotropic-equivalent total kinetic energy $E_{k,{\rm iso}}$; (2) CSM density, either for an ISM-like ($n$ constant) or a wind-like CSM ($\rho_{CSM} = \dot{M}/(4\pi R^2 v_w$) produced by a constant $\dot{M}$; (3) microphysical shock parameters $\epsilon_e$ and $\epsilon_B$; (4) jet opening angle $\theta_j$; (5) observer angle with respect to the jet axis $\theta_{\rm obs}$.
We fix the power-law index of the shocked electron energy distribution for a typical value in the range $p=2$--$3$, as derived from GRB afterglow modeling  (e.g., \citealt{Curran10,Wang15}) and we generate a model for a range of $\dot{M}$ for an assumed wind velocity of $v_w = 1000$~km s$^{-1}$.

\begin{figure*} %[ht]%[htbp]
\begin{center}
\scalebox{1.}
{\includegraphics[width=0.72\textwidth]{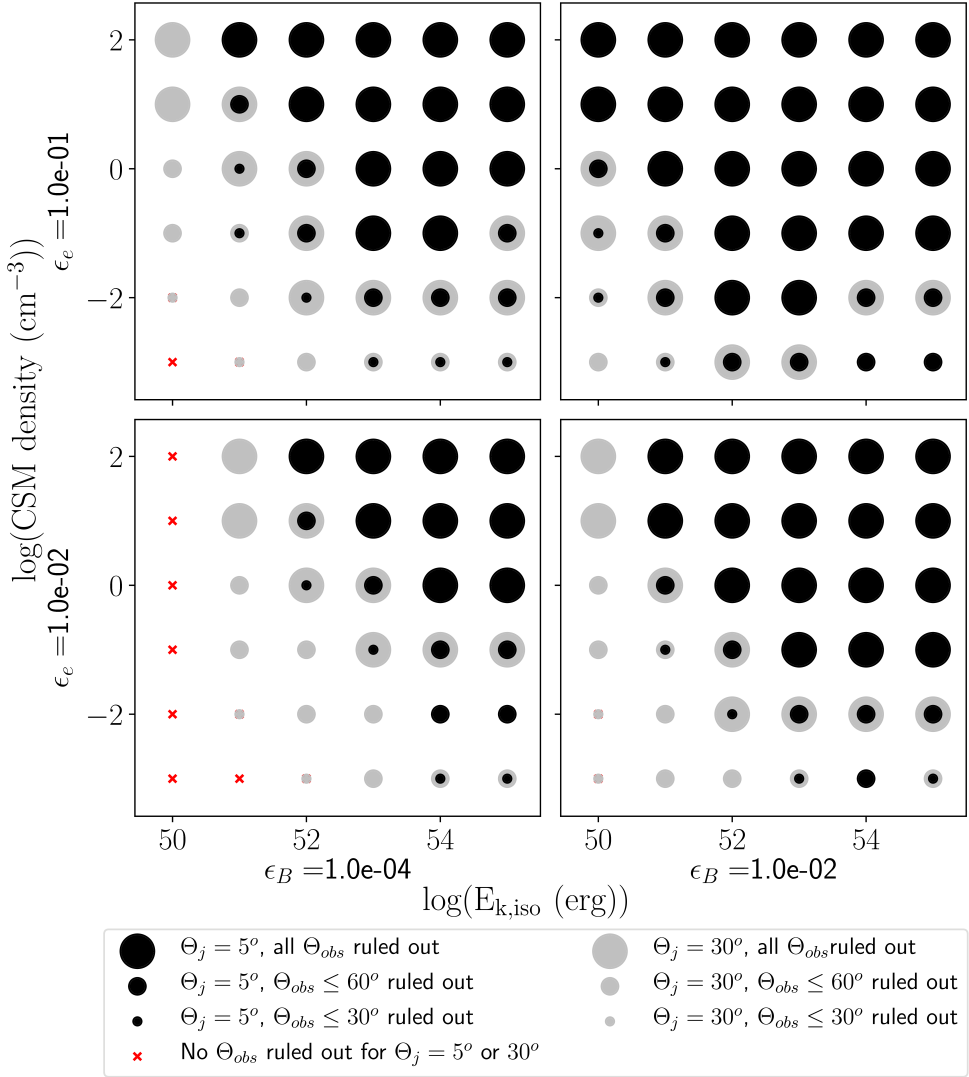}}
\caption{
Constraints on jetted outflows in an ISM-like density profile in the CSM, based on the VLA upper limits of SN\,2014ad and hydrodynamic simulations with {\sc boxfit}(v2) code (Sect.~\ref{sec:constraints}).
Black circles represents jet opening angles of $\theta_j = 5^{\circ}$, whereas gray circle represents jet opening angles of $\theta_j = 30^{\circ}$.
The symbol size indicates the observer angle ($\theta_{obs}$) out to which we can rule out the corresponding jet, with larger symbols corresponding to larger $\theta_{obs}$.
Red crosses indicate that we cannot rule out an off-axis relativistic jet with the given parameters in SN\,2014ad.
The top (bottom) panels are $\epsilon_e = 0.1$ ($\epsilon_e = 0.01$), and the left (right) panels are $\epsilon_B = 0.0001$ ($\epsilon_B = 0.01$). 
\label{fig:grid_boxfit_ism}
}
\end{center}
\end{figure*}

\begin{figure*} %[ht]%[htbp] 
\begin{center}
\scalebox{1.}
{\includegraphics[width=0.72\textwidth]{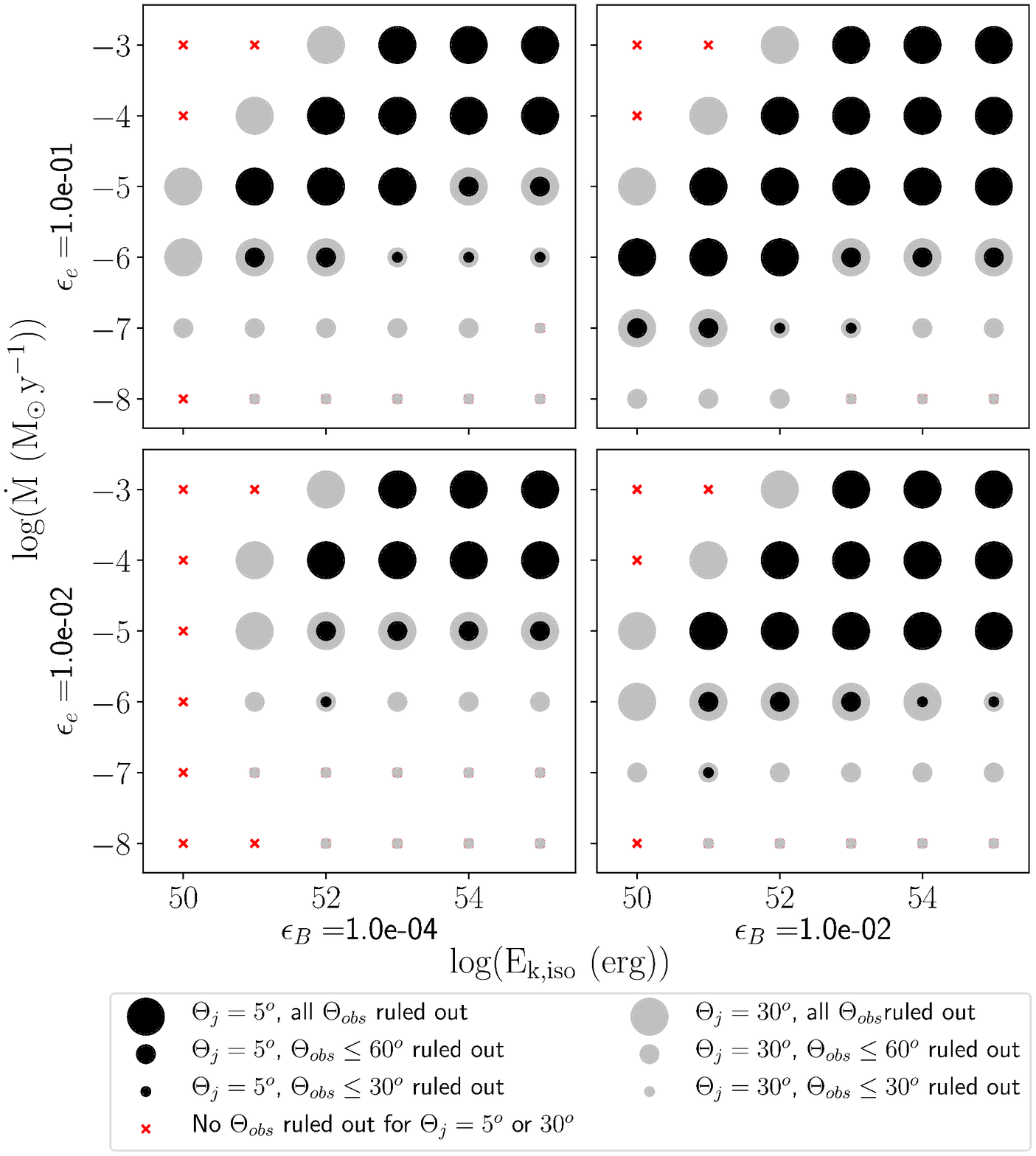}}
\caption{
Constraints on jetted outflows in a wind-like density profile in the CSM ($\rho \propto r^{-2}$), based on the VLA upper limits of SN\,2014ad and hydrodynamic simulations with {\sc boxfit}(v2) code (Sect.~\ref{sec:constraints}).
See the caption of Figure~\ref{fig:grid_boxfit_ism} for a full description of the symbols.
\label{fig:grid_boxfit_wind}
}
\end{center}
\end{figure*}

We explored a grid of parameters, specifically: $10^{-3}$ cm$^{-3} \leq n \leq 10^2$ cm$^{-3}$; $10^{-8}$ $M_\odot$ yr$^{-1} \leq \dot{M} \leq 10^{-3}$ $M_\odot$ yr$^{-1}$.
Two different jet opening angles were used, which encompass representative measured values for other GRBs: $\theta_j = 5^{\circ}$ and $30^{\circ}$.
We considered three observer angles ($\theta_{obs} = 30^{\circ}$, $60^{\circ}$, and $90^{\circ}$) and isotropic-equivalent kinetic energies in the range $10^{50}$~erg $\leq$ $E_{k,{\rm iso}} \leq 10^{55}$~erg. These ranges describe the typical parameters derived from accurate broadband modeling of GRB afterglows (e.g., \citealt{Schulze11,Laskar13,Perley14,Laskar16}).
Moreover, in this analysis we discuss the results for $\epsilon_e = 0.1$ and $\epsilon_B = 0.01$, but for completeness we show the results for other typical values in Figures~\ref{fig:grid_boxfit_ism} and \ref{fig:grid_boxfit_wind}. We find that our radio limits are consistent with the expected emission from off-axis ($\theta_{\rm obs} \geq 60^{\circ}$) narrow ($\theta_j = 5^{\circ}$) jets expanding in a low-density CSM environment with $\dot{M} \la 10^{-5}$~M$_{\odot}$ yr$^{-1}$ that are typical of BL-Ic SNe and GRBs. The allowed beaming-corrected kinetic energy values are $E_k \le 4 \times 10^{49}$~erg.

%-------------------------------------------
\section{Discussion}
\label{sec:disc}
%-------------------------------------------
Here we put our results on the environment and on the energetics of SN\,2014ad into the broader context of nearby ($z\le 0.2$) BL-Ic SNe with or without an associated GRB.

%-------------------------------------------
\subsection{Constraints on uncollimated outflows in SN\,2014ad}
\label{subsec:disc_ordinary}
%-------------------------------------------

In the case of sub-relativistic and nearly isotropic ejecta (Sect.~\ref{subsec:ssamod}) expanding in a wind-like CSM, assuming equipartition ($\epsilon_e = \epsilon_B = 1/3$),  Figure~\ref{fig:grid_nr_sn} shows that the combination of VLA + XRT data constrains the fast-ejecta kinetic energy to $E \la 10^{45}$~erg for $\dot{M}\la 10^{-6}~M_\odot$\,yr$^{-1}$ and to $E \la 10^{46}$~erg for $\dot{M}\la 10^{-4}M_\odot$\,yr$^{-1}$.
These very deep constraints rule out outflows with properties similar to (i) relativistic SNe, such as SN\,2009bb \citep{Soderberg10a} and SN\,2012ap \citep{Chakraborti15}, for which no GRB counterpart was detected, and (ii) SN\,1998bw, a prototypical GRB-SN associated with a low-luminosity GRB, propagating into a similar environment (Figure~\ref{fig:grid_nr_sn}). Our limits also point to very low density environments, consistent with previous findings that BL-Ic SNe favor low-density media (e.g., see Fig.~5 from \citealt{Margutti18b}), as was also the case for SN\,2002ap \citep{Berger02} and SN\,2010ay \citep{Sanders12b}.

%-------------------------------------------
\subsection{Is SN\,2014ad associated with an off-axis GRB-like jet?}
\label{subsec:disc_offaxis}
%-------------------------------------------

Our VLA radio observations place stringent constraints on off-axis relativistic jets expanding into an ISM-like (Figure~\ref{fig:grid_boxfit_ism}) and a wind-like CSM (Figure~\ref{fig:grid_boxfit_wind}), respectively (Sect.~\ref{sec:constraints}).  
First, we consider the case of a wind-like CSM  and a highly collimated jet with $\theta_j = 5^{\circ}$ (as is typical for cosmological GRBs) viewed off-axis, for $\epsilon_e = 0.1$ and $\epsilon_B = 0.01$ (top right panel, Figure \ref{fig:grid_boxfit_wind}).
These off-axis narrow jets are ruled out regardless of the observer angle for $\dot{M} \ga 10^{-5}$~M$_{\odot}$ yr$^{-1}$ and $E_{k,iso} \ga 10^{52}$~erg (typical value for a GRB).
Hence, GRB-like jets expanding either in a low-density CSM typical of BL-Ic SNe ($\dot{M} \la 10^{-5}$ -- $10^{-6}$~M$_{\odot}$ yr$^{-1}$ in Table~1 of \citealt{Smith14}; see also \citealt{LiChevalier99} and \citealt{Soderberg06d}) or in typical GRB environments ($10^{-7} \la \dot{M} \la 10^{-5}$~M$_{\odot}$ yr$^{-1}$; \citealt{Laskar14,Laskar15}) cannot be ruled out.

In the case of off-axis jets with larger opening angle $\theta_j = 30^{\circ}$, for $\epsilon_e = 0.1$ and $\epsilon_B = 0.01$ (top right panel, Figure \ref{fig:grid_boxfit_wind}), we obtain stronger constraints, due to their larger jet energy.
Specifically, regardless of the observer angle, we can rule out scenarios where $\dot{M} \ga 10^{-6}$~M$_{\odot}$ yr$^{-1}$ and $E_{k,{\rm iso}} \ga 10^{52}$~erg.
Mass-loss rates typically found in the winds of WR stars ($\dot{M} \la 10^{-5}$ -- $10^{-6}$~M$_{\odot}$ yr$^{-1}$; \citealt{Smith14}) are mostly ruled out.
In the case of wide ($\theta_j = 30^{\circ}$), slightly off-axis ($\theta_{obs}\le30^{\circ}$) jets, for $\epsilon_e = 0.1$ and $\epsilon_B = 0.01$ (top right panel, Figure \ref{fig:grid_boxfit_wind}), we can rule out the combination of $\dot{M} \ga 10^{-8}$~M$_{\odot}$ yr$^{-1}$ and $E_{k,{\rm iso}} \ga 10^{51}$~erg.
Assuming a progenitor wind velocity of $1000$~km s$^{-1}$, all the CSM profiles of all the detected SNe Ibc and most of the GRBs detected to date are rejected (see Figure 5 in \citealt{Coppejans18}).
We also report the results for a jet propagating into an ISM-like CSM, as the modeling of GRB afterglows often indicates an ISM environment as opposed to a wind-like density profile (e.g., \citealt{Laskar14,laskar18}).
For $\epsilon_e = 0.1$ and $\epsilon_B = 0.01$ (top right panel, Figure \ref{fig:grid_boxfit_ism}), highly collimated jets with $\theta_j = 5^{\circ}$ are ruled out regardless of the observer angle for $n \ga 10$~cm$^{-3}$ and $E_{k,{\rm iso}} \ga 10^{50}$~erg, or for $n \ga 10^{-1}$~cm$^{-3}$ and $E_{k,{\rm iso}} \ga 10^{52}$~erg. A jet with $\theta_j = 30^{\circ}$ is ruled out for $n \ga 10^{-1}$~cm$^{-3}$ and $E_{k,{\rm iso}} \ga 10^{50}$~erg.
We obtain deeper constraints for jets with $\theta_{obs} < 60^{\circ}$: for $\theta_j = 5^{\circ}$ and $\theta_{obs} = 60^{\circ}$ a jet is ruled out for $n \ga 10^{-3}$~cm$^{-3}$ and $E_{k,{\rm iso}} \ga 10^{52}$~erg.
Hence, GRB-like jets expanding in a ISM-like medium with $n \la 10^{-2}$~cm$^{-3}$ and $E_{k,{\rm iso}} \la 10^{50}$~erg cannot be ruled out: these densities are compatible with those of some GRBs ($10^{-5} \la n \la 10^3$~cm$^{-3}$; e.g.,  \citealt{Laskar14,Laskar15}).

We conclude that we cannot rule out the case of an off-axis ($\theta_{obs} \ga 30^{\circ}$), narrow ($\theta_j = 5^{\circ}$) GRB-like jet ploughing through low-density CSM typical of BL-Ic SNe and GRBs; this scenario allows for beaming-corrected kinetic energies $E_{k,iso} \la 10^{52}$~erg in environments sculpted by $\dot{M} \la 10^{-6}$~M$_{\odot}$ yr$^{-1}$.

%-------------------------------------------
\subsection{Constraining the $E_k(\Gamma\beta)$ distribution of the ejecta of SN\,2014ad}
\label{subsec:disc_others}
%-------------------------------------------
Compared with BL-Ic GRB-less SNe, GRB-SNe seemed to show (i) high mass of $^{56}$Ni synthetized in the SN explosion, (ii) higher degree of asphericity in the SN explosion, and (iii) low metallicity of the SN environment (e.g., \citealt{Cano13}).
However, \citet{Taddia19} recently showed that the distributions of these observables for the two classes of BL-Ic SNe are still compatible within uncertainties.
Another way to investigate the differences between the two classes is offered by the slope $x$ of the kinetic energy profile ($E_k$) as a function of the ejecta four-velocity ($\Gamma \beta$), parametrized as $E_k\propto (\Gamma\beta)^x$. What is more, this may help to reveal the nature of the explosion (see Fig.~2, \citealt{Margutti14b}).
Steep profiles ($x \la -2.4$) indicate a short-lived central engine, and hence an ordinary Ibc SN \citep{Lazzati12}; flat profiles ($x \ga -2.4$) indicate the presence of a mildly short-lived central engine, and hence a possible GRB-SN \citep{Margutti13b}; very flat profiles ($x = -0.4$) are typical of ordinary GRBs in the decelerating Blandford-McKee phase \citep{BM76}, whereas very steep profiles ($x = -5.2$) are characteristic of a pure hydrodynamical spherical explosion \citep{Tan01}.

\begin{figure} %[htbp] 
\centering
\includegraphics[width=8.8cm]{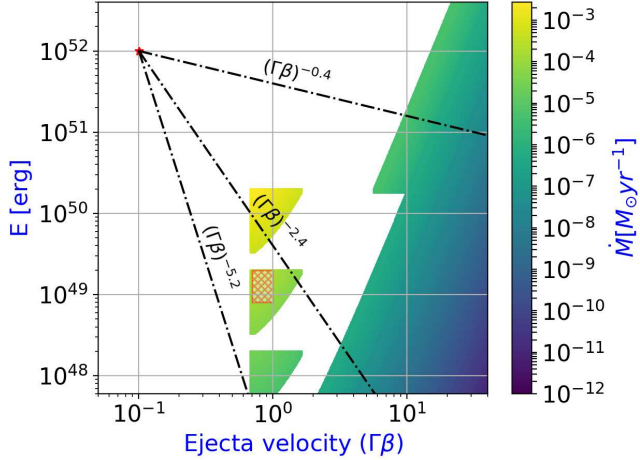}
\caption{
Region of the beaming-corrected energy $E_k$ -- ejecta velocity $\Gamma \beta$ (with $\Gamma$ estimated at $t = 1$~d) space allowed by our upper limits of SN\,2014ad (in wind-like CSM for relativistic regime).
The color scale shows the allowed progenitor mass-loss rate $\dot{M}$.
The dashdot lines indicate the slope $x$ of the kinetic energy profile.
The orange hatched area indicates the region of relativistic SNe, where the cocoon emission might be observable \citep{DeColle18a,DeColle18b}. \\
\label{fig:grid_ekgammabeta}
}
\end{figure}

For SN\,2014ad we explored a grid of parameters in the $E_k$ -- $\Gamma\beta$ space.
$\Gamma$ is calculated at $t = 1$~d applying the standard formulation of the fireball dynamics with expansion in a wind-like CSM (e.g., \citealt{ChevalierLi00})
\begin{equation}
\Gamma_{(t = 1\,d)} \sim 18.7 \left(\frac{E_{k,{\rm iso}}}{10^{54}erg}\right)^{1/4} \left(\frac{A_*}{0.1}\right)^{-1/4}\;,
\label{eq:gamma_wind}
\end{equation}
where $A_*$ is the circumstellar density, defined with respect to progenitor mass-loss rate $\dot{M}$ and wind velocity $v_w$ as:
\begin{equation}
A_*\ =\ \left(\frac{\dot{M}}{10^{-5}M_\odot \; {\rm yr}^{-1}}\right) \left(\frac{v_w}{1000 \; {\rm km\ s}^{-1}}\right)\;.
\label{eq:a_wind}
\end{equation}
The allowed regions are derived through the conditions described in Sect.~\ref{subsec:disc_offaxis} for the case of a highly collimated jet with $\theta_j = 5^{\circ}$ (as typical for cosmological GRBs) viewed off-axis in a wind-like CSM (Figure~\ref{fig:grid_boxfit_wind}; top right panel).
Figure~\ref{fig:grid_ekgammabeta} shows the allowed region of the beaming-corrected energy $E_k$ -- ejecta velocity $\Gamma \beta$ space (in the relativistic regime).
Relativistic jets for SN\,2014ad are possible for progenitors with very low densities ($\dot{M} \la 10^{-7}$~M$_{\odot}$ yr$^{-1}$); for example, a faster-moving ejecta (with a beaming-corrected energy $E_k \sim 10^{51}$~erg) ploughing through a wind-like CSM with a very low density $\dot{M} \sim 10^{-7}$~M$_{\odot}$ yr$^{-1}$ has $\Gamma \beta \sim 24$ (at $t = 1$~d), compatible with the flat profile ($x = -0.4$) of ordinary GRBs.
The lack of any associated GRB suggests a possible off-axis GRB propagating in a wind-like CSM with a very low density ($\dot{M} \la 10^{-7}$~M$_{\odot}$ yr$^{-1}$).

%-----------------------------------------------------
\subsection{Constraints on Cocoon emission in SN\,2014ad}
\label{subsec:cocoon}
%-----------------------------------------------------

The interaction between the jet emission and the outer layers of the progenitor star causes the swelling of the outer envelope of the jet, called cocoon.
The recent broadband spectroscopic analysis of \citet{Izzo19} of a BL-Ic GRB-SN (SN\,2017iuk/GRB\,171205A) shows the first direct evidence for the cocoon emission.
This cocoon is characterized by a very high expansion velocity ($\sim 0.3\,c$) and probably originates from the energy injection of a mildly-relativistic GRB jet.
This discovery could explain the lack of GRBs observed in association with some BL-Ic SNe: the jet, since it transfers a significant part of its total energy to the cocoon, produces the typical GRB emission only if it manages to completely pierce the star photosphere.
This conclusion is in agreement with the analysis of \citet{DeColle18a,DeColle18b}: they show that the radio emission observed in relativistic SNe can be explained as synchrotron emission from the cocoon created by an off-axis GRB jet (either failed or successful), that propagates through the progenitor star.
Figure~\ref{fig:grid_ekgammabeta} shows the allowed region (red hatched area) for relativistic SNe, where the cocoon emission in principle might be observable: even if the radio emission from SN\,2014ad is much fainter than SN\,2009bb and SN\,2012ap (Figure~\ref{fig:radioworld}), this region is compatible with $E_k$ of the fast ejecta for a SN\,2014ad progenitor with mildly-low densities ($\dot{M} \sim 10^{-5}$~M$_{\odot}$ yr$^{-1}$).
\citet{DeColle18a} suggest that, in the off-axis GRB scenario, the cocoon synchrotron emission at radio frequencies dominates (i) always for failed GRB/cocoon or weak GRB observed off-axis, or (ii) only at early times for energetic off-axis jets with late-times peaks (timescale of years).

A more quantitative discussion of the cocoon emission for SN\,2014ad is beyond the scope of the present investigation.

%-------------------------------------------
\section{Conclusions}
\label{sec:conc}

%-------------------------------------------

We present deep X-ray and radio limits of the BL-Ic SN\,2014ad.
Radio and X-ray observations are crucial for probing the fastest moving ejecta in the explosion, as the optical emission is produced by the slow-moving ejecta.
Previous studies of this source showed that it has a number of properties that, taken together, suggest a possible GRB counterpart. These include a large bulk energy $E_k$ of the slow ejecta, the asphericity in the explosion and ejecta velocity, the large inferred Nickel mass, and the low progenitor mass-loss rate $\dot{M}$.
Consequently, we investigated two different physical scenarios for SN\,2014ad:
(i) a sub-relativistic, nearly isotropic explosion of an ordinary BL-Ic SN in a wind-like CSM (Sect.~\ref{subsec:ssamod});
(ii) an off-axis relativistic jet (Sect.~\ref{sec:constraints}).
These models place strong constraints on the total energy of the fast ejecta ($E$), the progenitor mass loss rate ($\dot{M}$), the jet opening angle ($\theta_j$) and the observer angle ($\theta_{obs}$).
We obtained the following results:
\begin{itemize}
\item Assuming that the dominant source of X-ray emission at early times is IC emission from the upscattering of optical photospheric photons into the X-ray band by relativistic electrons at the shock front (Sect.~\ref{sec:IC}), we infer $\dot{M} < 5\times10^{-5}$~M$_{\sun}$ yr$^{-1}$, for a wind velocity $v_w = 1000$~km s$^{-1}$ for a spherical outflow.
\item If SN\,2014ad launched a sub-relativistic and isotropic outflow (Sect.~\ref{subsec:ssamod}), assuming equipartition ($\epsilon_e = \epsilon_B = 0.33$) we derive limits of $E \la 10^{45}$~erg for $\dot{M}\la 10^{-6}~M_\odot$\,yr$^{-1}$ and $E \la 10^{46}$~erg for $\dot{M}\la 10^{-4}M_\odot$\,yr$^{-1}$. 
These deep constraints rule out outflows with properties similar to (i) relativistic SN\,2009bb and SN\,2012ap, for which no associated GRB was reported, and (ii) SN\,1998bw, a prototypical GRB-SN, propagating into a similar environment.
$E$ and $\dot{M}$ of the kind seen in the GRB-less SN\,2002ap and SN\,2010ay, which are characterized by a modest energy budget in the fast ejecta, are not ruled out.  
\item If SN\,2014ad launched a relativistic jet, we (i) rule out collimated on-axis jets of the kind detected in GRBs, (ii) put strong constraints on the energies and CSM densities for an off-axis jet (Figure~\ref{fig:grid_boxfit_ism} and \ref{fig:grid_boxfit_wind}).
We cannot rule out an off-axis GRB in very low-density CSM environments (e.g., $\theta_{\rm obs} \ga 30^{\circ}$, $\theta_j = 5^\circ$, in a CSM sculpted by $\dot{M} \la 10^{-6}$~M$_{\odot}$ yr$^{-1}$, typical of BL-Ic SNe and GRBs).
Moreover, we cannot reject the possibility of a radio synchrotron emission dominated by the cocoon created by a GRB jet viewed off axis, that propagates through the stellar progenitor, as expected for relativistic SNe.
\end{itemize}
With our analysis of the off-axis jet scenario we have demonstrated that it is not possible to rule out off-axis jets expanding into low-density environments (as previously found by \citealt{Bietenholz14VLBI} for other SNe). For SN\,2014ad we find  $\dot{M} \la 10^{-6}$~M$_{\odot}$ yr$^{-1}$ (Figure \ref{fig:grid_ekgammabeta}).
\emph{If} SN\,2014ad  was indeed powered by an off-axis relativistic jet, our X-ray and radio observations imply extremely low environment densities and energies coupled to jet (unless the jet was far off-axis).

Deep radio and X-ray observations at early \emph{and} at late times of a large sample of nearby BL-Ic SNe will clarify if relativistic jets are ubiquitous in BL-Ic SNe.

%%%%%%%%%%%%%%%%%%%%%
\acknowledgments
We thank D.~K.~Sahu for kindly sharing their bolometric light curves.
M.M. thanks M.~Orienti and E.~Egron for their precious suggestions about VLA data reduction and Bath University for the hospitality during the final stages of this work.
We acknowledge University of Ferrara for use of the local HPC facility co-funded by the ``Large-Scale Facilities 2010'' project (grant 7746/2011).
We thank University of Ferrara and INFN--Ferrara for the access to the COKA GPU cluster.
This research was supported in part through the computational resources and staff contributions provided for the Quest high performance computing facility at Northwestern University which is jointly supported by the Office of the Provost, the Office for Research, and Northwestern University Information Technology.
We gratefully acknowledge Piero Rosati for granting us usage of proprietary HPC facility.
Development of the {\sc boxfit} code was supported in part by NASA through grant NNX10AF62G issued through the Astrophysics Theory Program and by the NSF through grant AST-1009863. 
Simulations for {\sc boxfit}v2 have been carried out in part on the computing facilities of the Computational Center for Particle and Astrophysics of the research cooperation ``Excellence Cluster Universe'' in Garching, Germany. Support for this work was provided by Universit\`a di Ferrara through grant FIR~2018 ``A Broad-band study of Cosmic Gamma-Ray Burst Prompt and Afterglow Emission" (PI Guidorzi).
The National Radio Astronomy Observatory is a facility of the National Science Foundation operated under cooperative agreement by Associated Universities, Inc..
%%%%%%%%%%%%%%%%%%%%%%%%%%%%%%%%%%%%

\bibliographystyle{apj}
\bibliography{alles_grbs}   % name your BibTeX data base

%\begin{thebibliography}{}

%\end{thebibliography}

\end{document}